\documentclass[aps,prb,twocolumn,showpacs]{revtex4}
\usepackage{graphics,amsmath}
\usepackage{tabularx,graphicx}
\usepackage[latin1]{inputenc}
\usepackage{epsfig}

\newcommand{\dxa}{\partial_{x_1}}
\newcommand{\dxb}{\partial_{x_2}}
\newcommand{\dya}{\partial_{y_1}}
\newcommand{\dyb}{\partial_{y_2}}

\newcommand{\dr}{\partial_r}
\newcommand{\dphi}{\partial_{\phi}}

\newcommand{\vintrr}{v ( r )}

\begin{document}
\title{Two-body problem in graphene}
\author{J. Sabio$^{1,2}$}
\author{F. Sols$^2$}
\author{F. Guinea$^1$}
\affiliation{$^1$Instituto de Ciencia de Materiales de Madrid
(CSIC), Sor Juana In\'es de la Cruz 3, E-28049 Madrid, Spain \\ $^2$
Departamento de F{\'\i}sica de Materiales, Universidad
  Complutense de Madrid, 28040 Madrid, Spain.}

\begin{abstract}
We study the problem of two Dirac particles interacting through
non-relativistic potentials and confined to a two-dimensional sheet,
which is the relevant case for graphene layers. The two-body problem cannot
be mapped into that of a single particle, due to the non-trivial coupling
between the center-of-mass and the relative coordinates, even in the
presence of central potentials. We focus on the case of zero total
momentum, which is equivalent to that of a single particle in a Sutherland lattice. We show that zero-energy states induce striking new features such as discontinuities in the relative wave function, for particles interacting through a step potential, and a concentration of relative density near the classical turning point, if particles interact via a Coulomb potential. In the latter case we also find that the two-body system becomes unstable above a critical coupling. These phenomena may have bearing on the nature of strong coupling phases in graphene.
\end{abstract}
 \maketitle
\section{Introduction}

The problem of interactions in graphene layers was a subject  of
research even before its isolation and characterization in the
laboratory \cite{Netal04, Netal05}. It is peculiar due to its
two-dimensional nature and to the honeycomb lattice structure into
which ions are arranged. In the low-energy limit, the electronic
properties are described by a Dirac-like equation for massless and
chiral electrons \cite{NGPNG09}. Undoped graphene has a vanishing
density of states at the Fermi level and therefore a diverging
screening length, so interactions are expected to yield a more
singular behavior than in a conventional Fermi liquid picture,
already well established in metals and electron gases \cite{PN66,
GV05}. In fact, a weak-coupling scaling analysis was early performed
\cite{Gon94,GGV99}, shedding light on the role of electron-electron
interactions mediated by the Coulomb potential: the resulting
divergences in perturbation theory, once handled conveniently, turn
out to have a small effect on the low energy properties of
electrons, as they are marginally irrelevant in the renormalization
group sense. However, that analysis does not exclude the possibility
of phases with broken symmetry, where interactions play a major
role, when the dimensionless coupling $g \equiv e^2/\epsilon \hbar
v_F$ is of order unity or larger, $e$ being the electronic charge,
$v_F$ the Fermi velocity, and $\epsilon$ the dielectric constant of
the environment in which graphene is embedded \cite{K01, GGMS02,
KL04}. This is expected to be the relevant case for graphene layers
in vacuum, where $\epsilon = 1$ and $g \simeq 2.16$. Along that
direction, several works in the recent literature are pointing out
the existence of exotic strong coupling phases in graphene layers,
where pairing of electron and holes could give rise to an excitonic
state, where a gap would be opened rendering the system insulating
\cite{K01, Her08, DL09}.

On the other hand, it is common wisdom in the field of strongly
correlated systems, that the study of the interaction between two
particles can provide important insights on the many-body physics.
The relevance of this kind of studies in graphene has already been
shown when addressing the Coulomb impurity problem \cite{PNN07,
Shy07, Shy07y2, N07}. Here, a critical value of the coupling marks the breakdown
of the Dirac vacuum, whose study requires consideration of the whole
many-body problem. As pointed out recently \cite{GGG09, Wang09},
there could be a relation between this instability and the formation
of an excitonic condensate in the strongly-coupled many-body
problem. Thus two-particle physics seems to underlie many features
of the full many-body problem.

In this paper, we address the problem of two interacting Dirac
electrons in two spatial dimensions, mediated by non-relativistic
central potentials. This feature makes the problem different from
the already well addressed fully relativistic problem. One of our
goals is to shed light on the relation between the two-body problem
and the many-body instabilities, so we pay special attention to the
case of the bare Coulomb potential. However, the general problem
happens to show peculiarities that make its separate study
worthwhile. Importantly, the two-body problem cannot be mapped
exactly into the one-body Coulomb impurity problem on the same lattice.
In fact, we will
show that the two-body problem presents remarkable differences, the
most important being the singular role played by localized zero-energy
states at those points where the kinetic energy vanishes.
Interestingly, we find that, for zero center-of-mass momentum, the two-body problem in graphene is equivalent to that of a single particle in the lattice model proposed by Sutherland \cite{S86}, where both lattices are considered in their continuum limit. The presence of zero-energy states can induce non-analyticities in the relative wave function, giving rise to partial localization phenomena for the Coulomb interacting case. In order
to get a better understanding of the novel features, we address
first the case of two particles interacting via a step potential.

The paper is organized as follows: Section II presents some  general
features of the two-body problem. Section III addresses the case of
zero center-of-mass momentum, which is simpler to analyze and of
potential relevance to many-body instabilities. Sections IV and V
study the case of step and Coulomb potentials, respectively. Section
VI discusses some aspects of the case of arbitrary center-of-mass
momentum. Section VII is devoted to a discussion of the relation of
the present work to many-body phenomena. Finally, section VIII
summarizes the main conclusions. The paper includes three appendices
where some technical issues have been collected.

\section{General features}

As the main applications of this problem concern graphene  sheets,
we start with a formulation of the problem in terms of the continuum
theory of graphene electron motion, which is known to be described by
the Dirac equation. For a single particle, the wave function is a
two-component spinor characterized by the
quantum numbers of spin and valley, both with degeneracy two. For
zero magnetic field and scalar translationally invariant
interactions such as those which we will consider here, we can
neglect the effect of those extra degrees of freedom and concentrate
on the two-component spinor description. Later, we will discuss the
effect of the spin. The single particle Dirac equation reads:
\begin{eqnarray}
\left[ \begin{array}{cc} v(r)  &  -i\partial_x -\partial_y \\
-i\partial_x + \partial_y & v(r) \end{array} \right]\left[
\begin{array}{c} \Psi_A({\bf r}) \\ \Psi_B({\bf r}) \end{array}
\right] = E \left[ \begin{array}{c} \Psi_A({\bf r}) \\
\Psi_B({\bf r}) \end{array} \right]
\end{eqnarray}
where the pseudo-spin index $i = A,B$ refers to the  two
inequivalent sites within the unit cell of the honeycomb lattice.

Now let us consider the two particle problem. We  construct
two-particle wave functions from the tensor product of
single-particle ones, $\Psi_{ij}({\bf r}_1, {\bf r}_2) \equiv
\Psi_i({\bf r}_1)\otimes \Psi_j({\bf r}_2)$. This allows us to write
the Schr\"odinger equation for the interacting problem, that in the language of four-component spinors reads:
\begin{widetext}
\begin{equation}
\left[ \begin{array}{cccc} v(r) & -i\dxb -\dyb & -i\dxa -\dya &0
\\-i\dxb + \dyb & v(r)  &0 & -i\dxa -\dya\\ -i\dxa + \dya &0 & v(r)
& -i\dxb -\dyb \\ 0 & -i\dxa + \dya & -i\dxb + \dyb & v(r)
\end{array} \right] \left[\begin{array}{c} \Psi_{AA} ({\bf r}_1, {\bf r}_2)
\\ \Psi_{AB} ({\bf r}_1, {\bf r}_2) \\ \Psi_{BA} ({\bf r}_1,
{\bf r}_2) \\ \Psi_{BB} ({\bf r}_1, {\bf r}_2)
\end{array} \right] = E  \left[ \begin{array}{c}
\Psi_{AA} ({\bf r}_1, {\bf r}_2)
\\ \Psi_{AB} ({\bf r}_1, {\bf r}_2) \\ \Psi_{BA} ({\bf r}_1, {\bf r}_2)
\\ \Psi_{BB} ({\bf r}_1, {\bf r}_2)
\end{array} \right]
\end{equation}
\end{widetext}
As we are dealing with translationally invariant potentials,  we can
switch to the center-of-mass frame, defining the new coordinates
${\bf R} = \frac{{\bf r}_1 + {\bf r}_2}{2}$ and ${\bf r} = {\bf r}_1
- {\bf r}_2$. It is also convenient to apply the unitary
transformation
\begin{eqnarray}
\Psi_{1} = \Psi_{AA} \nonumber\\
\Psi_{2} = \frac{1}{\sqrt{2}}(\Psi_{AB} + \Psi_{BA}) \nonumber \\
\Psi_{3} = \frac{1}{\sqrt{2}}(\Psi_{AB} - \Psi_{BA}) \nonumber \\
\Psi_{4} = \Psi_{BB}
\end{eqnarray}
and use a plane wave ansatz for the center-of-mass part of the
wave function, $\Psi_i({\bf R}, {\bf r}) = e^{i {\bf K} \cdot
{\bf R}} \psi_i({\bf r})$. We arrive at the following eigenvalue
problem:
\begin{widetext}
\begin{equation}
\left[ \begin{array}{cccc} \vintrr  & \frac{1}{\sqrt{2}}K
e^{-i\theta_K} & \sqrt{2} e^{-i\phi}(i\dr + \frac{1}{r}\dphi) &0 \\
\frac{1}{\sqrt{2}}K e^{i\theta_K} & \vintrr & 0 &
\frac{1}{\sqrt{2}}K e^{-i\theta_K}
\\\sqrt{2}e^{i\phi}(i\dr - \frac{1}{r}\dphi) &  0 & v(r)
& -\sqrt{2}e^{-i\phi}(i\dr +\frac{1}{r}\dphi) \\ 0 &
\frac{1}{\sqrt{2}}K e^{i\theta_K} & -\sqrt{2} e^{i\phi}(i\dr -
\frac{1}{r}\dphi)  & v(r) \end{array} \right]
\left[\begin{array}{c} \psi_{1}
\\ \psi_{2} \\ \psi_{3} \\ \psi_{4} \end{array} \right] =
E  \left[ \begin{array}{c}  \psi_{1} \\ \psi_{2}   \\ \psi_{3} \\ \psi_{4}
\end{array} \right] \, ,
\label{Full_H}
\end{equation}
\end{widetext}
where $\theta_K \equiv \arctan (K_y/K_x)$ and polar
coordinates are used for the relative coordinate. As a first remark
on this equation, we notice that the center-of-mass coordinate does
not decouple from the relative one, even though the potential only
depends on the latter. This is a consequence of the chiral nature of
the electron carriers, where pseudo-spin and momentum are coupled.
This kind of coupling also prevents the Hamiltonian from commuting
with the relative angular momentum, thus frustrating a possible
decomposition of the problem in terms of partial waves.

\section{The case $K = 0$}

In order to gain insight into the many-body problem, the most
interesting case is that of zero total center-of-mass momentum. Then
the two particles have opposite momenta, like in the Cooper
channel in metals. Any pairing effect should be
particularly important in this energetically most favorable case. It
is also the simplest one, because it decouples the second component
$\psi_2(r)$ from the rest. In effect, the equation for this component reads
\begin{equation}
[v(r) - E]\psi_2(r) = 0
\label{zero-energy}
\end{equation}
whose solution is $\psi_2(r) = 0$ except at the particular  point
$v(r) = E$, if it exists. That point having measure zero, we will
ignore the $\psi_2$ component as physically irrelevant. However,
we will see later that, at the point where $v(r) = E$ is
satisfied, zero-energy states are responsible for important non-analyticities in the other components.

\begin{figure}
\begin{center}
\includegraphics[width=3in]{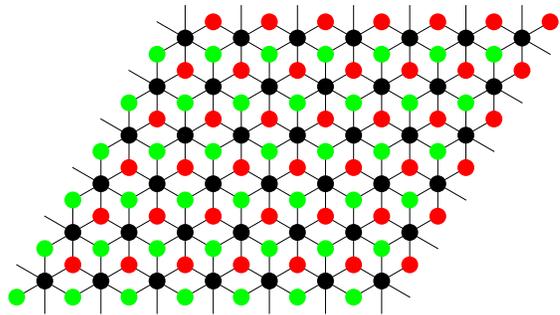}
\caption{Scheme of the lattice proposed by Sutherland in Ref. \onlinecite{S86}. The two-body problem in the low-energy sector of the honeycomb lattice, for $K = 0$, is mathematically equivalent to a single-particle problem in this lattice. Zero-energy states in the Sutherland lattice appear due to the existence of a flat band.}
\label{Sutherland}
\end{center}
\end{figure}

We are thus left with an effective three-component problem.
Remarkably, the $K = 0$ Hamiltonian commutes with the relative
angular momentum, so we can use the following ansatz for the
wave function:
 \begin{equation}
\left[ \begin{array}{c} \psi_1(r) \\ \psi_3(r) \\ \psi_4(r)
\end{array} \right] \equiv \left[ \begin{array}{c} e^{i(l-1)\phi}
\phi_1(r) \\ - \frac{i}{\sqrt{2}} e^{i l \phi}\phi_2(r) \\
e^{i(l+1)\phi}\phi_3(r) \end{array} \right] \label{three_comp}
\end{equation}
where the prefactors have been chosen for convenience. The labeling of the components in Eq. (\ref{three_comp}) has been
changed in order to accommodate it to the three-component case.
After using this ansatz, the system of equations reads:

\begin{equation}
\left[ \begin{array}{ccc} \vintrr - E & \dr + \frac{l}{r} &0
\\- 2 (\dr - \frac{l-1}{r}) & v(r) - E & 2 (\dr +\frac{l+1}{r}) \\
0 & -\dr + \frac{l}{r} & v(r) - E\end{array} \right]  \left[\begin{array}{c} \phi_{1}(r)
\\ \phi_{2}(r) \\ \phi_{3}(r) \end{array} \right] = 0
\label{final_H}
\end{equation}
It is interesting to note that these equations, as well as those directly derived from the full Hamiltonian for $K=0$,
Eq. (\ref{Full_H}),  can also be obtained as the continuum limit of a
one-particle lattice Hamiltonian, defined in a triangular lattice with three
sites in the unit cell, as initially considered by
Sutherland\cite{S86}. A scheme of this lattice is shown in Fig. \ref{Sutherland}.

Within this formulation, the case $l = 0$ is the most symmetric one:
 \begin{equation}
\psi(r) = \left[ \begin{array}{c} e^{-i \phi} \phi_1(r) \\ -
\frac{i}{\sqrt{2}} \phi_2(r) \\ e^{i \phi}\phi_3(r) \end{array} \right] \, .
\end{equation}
Henceforth, we will refer to it as the $s$-wave, and as we will see,
simple solutions can be obtained
for this case taking advantage of its symmetry.

\subsection{Symmetry properties}

Let us now analyze the symmetry properties of the $K = 0$ solutions.
In the original basis, the two-body wave functions read:
\begin{eqnarray}
\left[ \begin{array}{c}  \Psi_{AA} ({\bf r}_1, {\bf r}_2)
\\ \Psi_{AB} ({\bf r}_1, {\bf r}_2) \\ \Psi_{BA} ({\bf r}_1,
{\bf r}_2) \\ \Psi_{BB} ({\bf r}_1, {\bf r}_2)
\end{array} \right]_{K =0} = \left[ \begin{array}{c}  e^{i(l-1) \phi} \phi_1(r) \\
- \frac{i}{2} e^{i l \phi} \phi_2(r) \\  \frac{i}{2} e^{i l \phi}\phi_2(r)  \\
e^{i(l+1) \phi}\phi_3(r) \end{array} \right] \, ,
\end{eqnarray}
where the symmetric combination has been taken $\psi_2=0$, as argued above. This
wave function has a spinorial structure, due to the pseudo-spin of
the particles, and a spatial structure coupled to the former. The
symmetry properties under exchange of particles are studied by doing
the transformation:
\begin{eqnarray}
{\bf r}_1     \rightleftharpoons {\bf r}_2 \nonumber \\
\Psi_{AB} \rightarrow \Psi_{BA} \, .
\end{eqnarray}
The first transformation, for $K = 0$, translates into $\phi
\rightarrow \phi + \pi$. It follows inmediatly that wave functions
with $l$ even are antisymmetric under particle exchange,
while those with $l$ odd are symmetric. Hence, the $s$-wave is,
interestingly, antisymmetric. This somewhat counterintuitive result
reflects the role of the sublattice pseudo-spin in the orbital wave
function.

This has consequences on the total wave functions, once both spin
and valley degrees of freedom are also considered. Everywhere in
this article we assume that the two particles belong to the same
valley. Since their total wave function must be antisymmetric, the
following two families of solutions appear:
\begin{eqnarray}
\begin{array}{cc} \rm{(i)}  & \Psi_{K=0, l=\rm{odd}}({\bf r}_1,
{\bf r}_2) \otimes \frac{1}{\sqrt{2}}\left( |\uparrow \downarrow
\rangle - |\downarrow \uparrow \rangle \right) \\
\rm{(ii)} & \Psi_{K=0, l=\rm{even}}({\bf r}_1, {\bf r}_2) \otimes
|\uparrow \uparrow \rangle \\  & \Psi_{K=0, l=\rm{even}}({\bf r}_1,
{\bf r}_2) \otimes  |\downarrow \downarrow \rangle  \\ & \Psi_{K=0,
l=\rm{even}}({\bf r}_1, {\bf r}_2) \otimes \frac{1}{\sqrt{2}}\left(
|\uparrow \downarrow \rangle + |\downarrow \uparrow \rangle
\right)\end{array}
\end{eqnarray}
Therefore, the lowest angular-momentum channel ($l = 0$)
corresponds to a triplet spin-state, as opposed to what happens with
ordinary Schr\"odinger electrons.

\subsection{Other mathematical properties}

Equation (\ref{final_H}) comprises three coupled differential equations.
Adding the first and the third equations we may solve for $\phi _{2}$ in
terms of $\phi _{1}$ and $\phi _{3}$:
\begin{equation}
\phi _{2}=\frac{r}{2l}\varepsilon (r)(\phi _{1}+\phi _{3})~,  \label{phi-2}
\end{equation}%
where $\varepsilon (r)\equiv E-v(r)$. Substracting the same two equations,
we obtain

\begin{equation}
\partial _{r}\phi _{2}=\varepsilon (r)(\phi _{3}-\phi _{1})~.
\label{par-phi-2}
\end{equation}%
Differentiating Eq. (\ref{phi-2}) and relating the result to Eq. (\ref{par-phi-2}),
we obtain%
\begin{equation}
\partial _{r}[\varepsilon (\phi _{1}+\phi _{3})]=\frac{\varepsilon }{r}\left[
(2l-1)\phi _{3}-(2l+1)\phi _{1}\right] ~.  \label{system-1}
\end{equation}%
On the other hand, the second equation of system (\ref{final_H}) can be
rewritten as:%
\begin{equation}
\partial _{r}(\phi _{1}-\phi _{3})=\left( \frac{l-1}{r}-\frac{\varepsilon ^{2}r}{4l}%
\right) \phi _{1}+\left( \frac{l+1}{r}-\frac{\varepsilon ^{2}r}{4l}\right) \phi _{3}~.
\label{system-2}
\end{equation}%
Thus, system (\ref{final_H}) can be solved in principle by first solving for
$\phi _{1}$ and $\phi _{3}$ from (\ref{system-1}) and (\ref{system-2}) and
then obtaining $\phi _{2}$ from (\ref{phi-2}). The absolute values of $\phi
_{1}$ and $\phi _{2}$ should remain bounded as long as $\varepsilon (r)$ is
bounded. A problem may appear for $r\rightarrow 0$ in the case of the
Coulomb potential ($v(r)\sim r^{-1}$). In such a case, we will see that
regular solutions can be obtained analytically for $r\rightarrow 0$,
yielding in fact a useful starting point to initiate the numerical
integration.

Another important issue arises when $\varepsilon \rightarrow 0$, i.e., at those points where the kinetic energy vanishes, whenever $l >0$. For a smooth potential, a linear approximation of $\varepsilon(r)$ around the vanishing point $r_0$ holds, $\varepsilon(r) = \lambda x + {\mathcal O}(x^2)$, where $x \equiv r - r_0$. The differential equations (\ref{system-1}) and (\ref{system-2}) can be approximated around this point, yielding:
\begin{eqnarray}
\frac{d}{dx}\left[x (\phi_1 + \phi_3)\right] \simeq 0 \\
\frac{d}{dx}(\phi_1 - \phi_3) \simeq - \frac{1}{r_0}(\phi_1+\phi_3)
\end{eqnarray}
The solution for these equations reads $\phi_1 + \phi_3 \simeq - 2 C_2/ x$ and $\phi_1 - \phi_3 \simeq 2C_1+ (2 C_2/r_0)\log(x)$. We see that, for $l > 0$, a smooth potential will show non-analyticities close to $r_0$ in $\phi_1$ and $\phi_3$, while $\phi_2$ will remain continuous:
\begin{eqnarray}
\phi_1 &\simeq& C_1 + \frac{C_2}{r_0} \log(x) - \frac{C_2}{x} \nonumber\\
\phi_2 &\simeq& \frac{\lambda r_0}{2 l} C_2 \nonumber \\
\phi_3 &\simeq& - C_1 - \frac{C_2}{r_0} \log(x) - \frac{C_2}{x}
\end{eqnarray}
Notice, however, that these non-analyticites give a finite contribution to the probability  $\int dr \, r \,|\phi|^2$. Therefore they are physical solutions of the Dirac equation.

We end by noting that this singular behavior remains unaltered even in the pres-
ence of a small mass in the two-electron Dirac equation. 
Mathematically, this happens because the mass terms in the equations are
subleading in the short-distance expansion around the point $r_0$.

\section{Step potential}

\begin{figure}
\begin{center}
\includegraphics[width=3in]{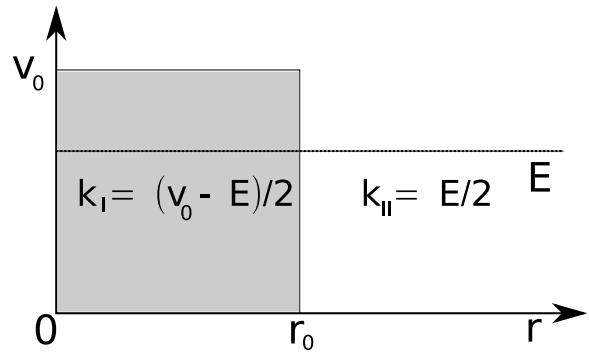}
\caption{Scattering of two particles interacting through a short range potential.}
\label{step}
\end{center}
\end{figure}

Some physical insight into the subtle properties of the interacting two-particle
problem can be obtained by considering the simpler
case of a step potential, which is typically
considered a good effective description of the more general class of
short-range potentials:
\begin{equation}
v(r) = \left\{ \begin{array}{lr} v_0 &  r<r_0\\ 0 & r>r_0
\end{array} \right. \label{step_v0}
\end{equation}
As usual, the procedure is to construct the solutions for each region and
eventually match them, as shown schematically in Fig. \ref{step}.

For arbitrary energy $E$, the solutions are given by Bessel
functions of the form:
\begin{equation}
\chi_1 = \left[ \begin{array}{c} a J_{l-1}(kr) \\ b J_{l}(kr)  \\ c
J_{l+1} (kr) \end{array} \right], \, \chi_2 = \left[
\begin{array}{c} a Y_{l-1}(kr) \\ b Y_{l}(kr) \\ c Y_{l+1} (kr)
\end{array} \right] \label{besselJ}
\end{equation}
where the coefficients and the eigenvalues are determined from the diagonalization of Eq. (\ref{final_H}), the result being
\begin{eqnarray}
E &=& v_0 + 2 k,\,\,\,\,  \left(\begin{array}{ccc}\frac{1}{2}, & 1, & \frac{1}{2} \end{array} \right)\\
E &=& v_0 - 2 k, \,\,\,\, \left(\begin{array}{ccc}\frac{1}{2}, & -1, & \frac{1}{2} \end{array} \right)\\
E &=& v_0, \,\,\,\, \frac{1}{\sqrt{2}}\left(\begin{array}{ccc}1, & 0, & -1 \end{array} \right)
\label{E-solutions}
\end{eqnarray}
the normalization being chosen to simplify the global wave function, Eq. (\ref{three_comp}). The first solution corresponds to two electrons located  in the
upper Dirac cone, while in the second solution the two electrons are
in the lower cone. The third solution describes the case where one
particle is in the upper cone and the other one in the lower cone.
Due to the zero total center-of-mass momentum, this solution has
zero total energy. Notice that, for fixed $E$, the relation of $k$
with the energy depends on the solution chosen. The third one is
valid for arbitrary $k$. Importantly, when $E = v_0$ there are other
zero-energy states that are also solutions of the two-particle Dirac
equation. They have the form
\begin{eqnarray}
\chi_3 = \frac{r^{\alpha}}{
[1 + (\frac{\alpha - l+1}{\alpha + l+ 1})^2]^{1/2}
}
\left(\begin{array}{ccc}1, & 0, & \frac{\alpha-l+1}{\alpha
+ l + 1} \end{array} \right) \, , \label{localized}
\end{eqnarray}
where $\alpha$ is a continuous parameter that can take any real
value. These polynomial solutions are in general non-physical, as
they cannot be properly normalized. However, they can be considered responsible for the non-analyticities shown to exist at those points where the kinetic energy vanishes (see section III.B). For the step potential, which is non-analytic itself,
the existence of zero-energy states induces discontinuities in the radial wave function, thus changing the usual matching
conditions. How such an anomalous behavior arises is explained in detail in Appendix A. Notice that in the
presence of a mass this kind of solutions still exist, since they involve
large derivatives within a narrow distance range.

\begin{figure}[t]
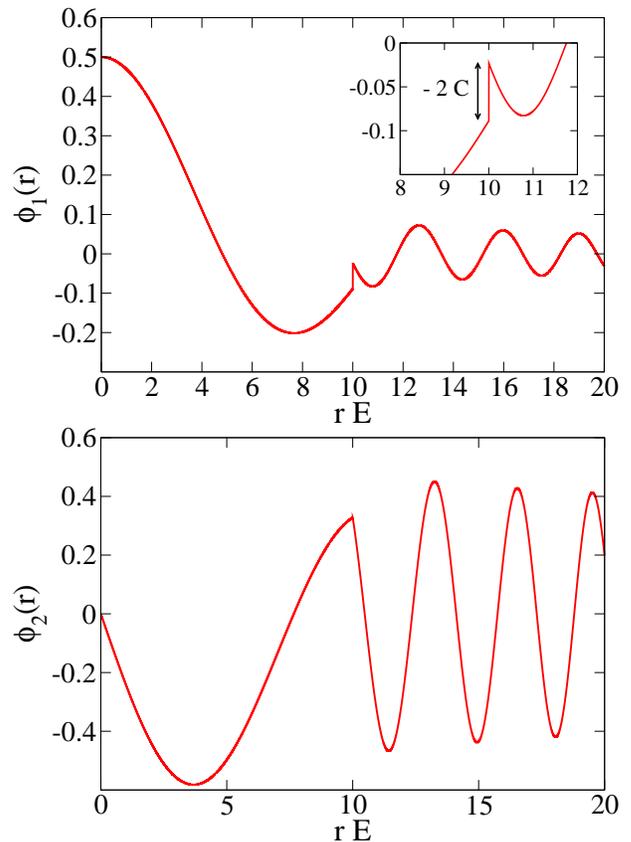

\begin{center}
\includegraphics*[width=0.45\textwidth]{fig3a.eps}
\includegraphics*[width=0.45\textwidth]{fig3b.eps}
\caption{Numerical solution of the step-potential for  $K = 0$
and $l = 1$ ($r_0 E = 10$, $E = v_0 /2$). Top: first component of
the radial wave function. The discontinuity induced by the zero
energy states arises naturally in the numerical solution. Bottom:
second component of the radial wave function. As also predicted by
the new matching conditions, the second component does not have a
d  iscontinuity} \label{step_numeric}
\end{center}
\end{figure}

We note that the
traditional criterion of imposing continuity of the wave functions
does not work here due to the insufficient number of matching
parameters.
To exemplify this issue, let us consider the situation shown  in
Fig. \ref{step}, where  $k_I = (v_0 - E)/2 > 0$  and $k_{II} = E /2>
0$. The solution for region I only includes Bessel functions of the
first kind, $J_l(k_I r)$, as those of the second kind ones are
singular at the origin. Hence $\phi_i^{I} \sim A_1 J_{l_i}(k_I r)$.
For region II both solutions must be considered, $\phi_i^{II} \sim
B_1 J_{l_i}(k_I r) + B_2 Y_{l_i}(k_{II} (r)$. Due to the freedom for
global normalization, only two relative values of the three
constants are relevant. Thus only two parameters remain to satisfy
the continuity of three equations, one for every component of the
wave function, leaving the problem overdetermined.

In Appendix A it is shown that, when zero-energy states  at the
matching point are taken into account, a third matching parameter
arises naturally which permits a discontinuity in the radial wave
function. Namely, we obtain
\begin{eqnarray}
\Delta \phi_1(r_0) \equiv \phi_1^{II}(r_0) - \phi_1^{I}(r_0) &=& - 2 C \nonumber \\
\Delta \phi_2(r_0) &=& 0 \nonumber \\
\Delta \phi_3(r_0) &=& - 2 C \, ,
\label{matching}
\end{eqnarray}
where $C$ is the extra parameter to determine. With the new
matching conditions, an exact solution of the $K = 0$
case becomes possible, as detailed
in Appendix B. It is also interesting to compute the solution
of the differential equations numerically, as shown in Fig.
\ref{step_numeric}. Here, the first two components of the radial
wave function are plotted. As predicted in equation
(\ref{matching}), the first component shows a discontinuity induced
by zero-energy states at the point $r = r_0$. The same consideration
applies to the third component (not shown) but not for the second
one.

It is also worth noting that, as expected, the $s$-wave shows  a
simpler behavior by virtue of its symmetric form. In this case,
inspection of Eq. (\ref{final_H}) shows $\phi_1(r) = - \phi_3(r)$
and the problem reduces to a two-component one. The matching
conditions reduce then to continuity and the $s$-wave problem
essentially behaves as that of the single particle. As a corollary, Eq.
(\ref{matching}) leads to $C=0$ in this case. Thus we may
state that the $l=0$ case has a structure similar to that of the
impurity one-body problem in graphene.

\section{Coulomb potential}

We now turn to the more relevant case of a long-range Coulomb
potential, $v(r) = g/r$, where $g = e^2/\epsilon \hbar v_F$ for
low-energy graphene electrons.  It is convenient to find a more
suitable form of Eq. (\ref{final_H}) in order to obtain analytical
solutions when possible. This is done by the usual procedure of
analyzing the short and long distance limits. At short distances
the wave function components have the form $\phi_i(r\rightarrow 0) \sim
r^{\gamma-1/2}$, with $\gamma^2 = \frac{1}{4}(1 + 4l^2- g^2)$. On
the other hand, the long-distance wave function behaves like a
plane wave of the form $\phi_i(r \rightarrow \infty) \sim
e^{\pm i E r/2}$. Hence, we can make the following
ansatz for the radial wave function:
\begin{equation}
\phi_i(\rho) = \rho^{\gamma-1/2} e^{- \rho/2} \hat{\phi_i}(\rho)
\end{equation}
where we have introduced the dimensionless radial complex  coordinate
$\rho = i E r$. By applying this transformation, Eq. (\ref{final_H})
becomes
\begin{widetext}
\begin{equation}
\left[ \begin{array}{ccc} i\rho + g & \rho \partial_{\rho} -
\frac{\rho}{2}+ \gamma + l - \frac{1}{2} &0
\\ -2(\rho \partial_{\rho} - \frac{\rho}{2}+ \gamma - l +
\frac{1}{2}) &i \rho + g&  2 (\rho \partial_{\rho} - \frac{\rho}{2}+
\gamma + l + \frac{1}{2})  \\ 0 & -\rho \partial_{\rho} + \frac{\rho}{2}-
\gamma + l + \frac{1}{2}  & i\rho + g \end{array} \right]
\left[\begin{array}{c} \hat{\phi}_{1}(\rho)
\\ \hat{\phi}_{2}(\rho) \\ \hat{\phi}_{3}(\rho) \end{array} \right] = 0\, .
\label{Coulomb_system}
\end{equation}
\end{widetext}
The general case is difficult to handle and only the $s$-wave
channel  admits an analytical solution, since it reduces to
an effective single particle problem. The details of this solution
are summarized in Appendix C. For general angular momentum $l$, Eq. (\ref{Coulomb_system}) must be solved numerically.

Before addressing the full solution, let us point out the remarkable
behavior of the wave functions at short distances. As we have seen,
it goes like a power law $r^{\gamma - 1/2}$, where $\gamma =
\frac{1}{2}\sqrt{1 + 4l^2- g^2}$. For $|g| < g_c \equiv \sqrt{1 + 4 l^2}$ only $\gamma > 0$ is acceptable. By contrast, for $|g| > g_c$, the $\gamma$ parameter becomes imaginary and the
wave function shows a pathological short-distance behavior, going like $r^{-1/2}[\cos(|\gamma|\log r) \pm i \sin(|\gamma| \log
r)]$. Thus the wave function oscillates dramatically towards the center. This kind
of behavior was already found in the Coulomb impurity problem, where
it was related to an instability of the wave function that could
signal the breakdown of the Dirac vacuum. For strong enough
couplings, the two-particle interaction would produce
electron-hole pairs from the vacuum, and a full quantum field-theoretical treatment
of the problem could be necessary. The consequences for the two-body problem of this effect have not been addressed in this paper, although from the study of the Coulomb impurity problem, we may expect a non-linear screening as a result of the reorganization of the many-body vacuum. \cite{Shy07y2}

\begin{figure}[t]
\begin{center}
\includegraphics*[angle=0, width = 0.45\textwidth]{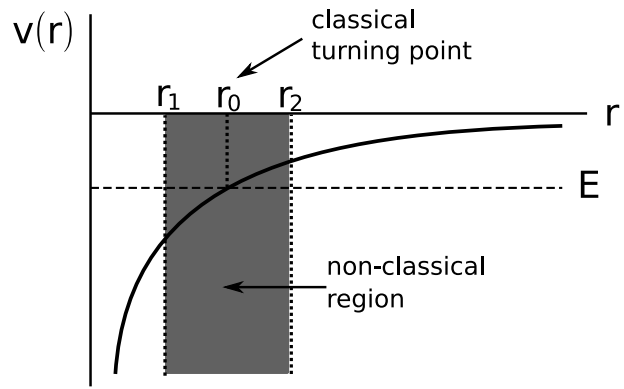}
\caption{Sketch of the Coulomb potential (attractive, in this case), including all the relevant lenght scales discussed in the text. $E$ is the total energy of the particle, $r_0$ is the classical turning point, and the region between $r_1$ and $r_2$ is the classically forbidden region.}
\label{clas}
\end{center}
\end{figure}

In order to gain further insight into the instability, a
semiclassical analysis like that performed in Ref. \onlinecite{Shy07}
can be applied. Our starting point is Eq. (\ref{final_H}) for the
Coulomb potential with the ansatz $\phi_i(r) =
\bar{\phi}_i \frac{e^{i 2 p_r r}}{\sqrt{r}}$, where $p_r$ is a
slowly varying function of $r$. This justifies the assumption
$\partial_r p_r \simeq 0$. Equation (\ref{final_H}) then
reads
\begin{equation}
\left[ \begin{array}{ccc} \frac{g}{r} - E  & i 2 p_r + \frac{l-1/2}{r} &0
\\-i 4 p_r + \frac{2l-1}{r} & \frac{g}{r} - E & i 4 p_r +\frac{2l+1}{r}  \\
0 & -i 2 p_r +\frac{l+1/2}{r} & \frac{g}{r} - E \end{array} \right]
\left[\begin{array}{c} \bar{\phi_{1}}
\\ \bar{\phi_{2}} \\ \bar{\phi_{3}}\end{array} \right] = 0 \, .
\end{equation}
The determinant vanishes if one of the following relations is satisfied:
\begin{eqnarray}
\frac{g}{r} = E \, , \\
p_r^2 = (\frac{g}{r}-E)^2 -
\frac{1}{r^2}(4l^2 + 1) \, .
\end{eqnarray}
The first equation defines a point, $r_0 \equiv g/E$,  where any
function $p_r$ gives a solution thanks to the existence of zero-energy states, as discussed in the previous section for the step
potential. Notice that this condition only is fulfilled for
repulsive electrons with positive energy or attractive electrons
with negative energy, two cases which are related through a symmetry
transformation. The second equation, on the other hand, defines a
non-classical region where $p_r^2 < 0$. The region is $r_1 < r <
r_2$, where $r_{1,2} = \frac{g}{E} \mp \frac{1}{E}\sqrt{4l^2 + 1}$.
Remarkably, we see that $r_1 < r_0 < r_2$, i.e. the point where
zero-energy states nucleate belongs to this classically forbidden region, as sketched in Fig. \ref{clas}.

\begin{figure}[tb]
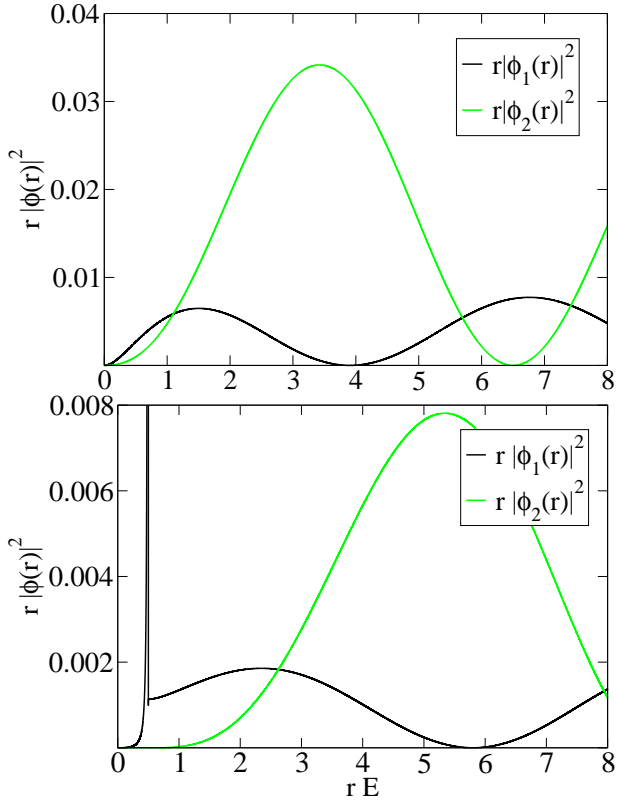

\begin{center}
\includegraphics*[angle=0, width = 0.45\textwidth]{fig5a.eps}
\includegraphics*[angle=0, width = 0.45\textwidth]{fig5b.eps}
\caption{Numerical solution of the radial wave function for the case of
Coulomb interaction and zero center-of-mass momentum. The chosen angular
momentum is $l = 1$. Top: wave functions for
$g = - 0.5$ and $E > 0$. Bottom: wave functions for $g = 0.5$ and $E
> 0$. Notice that, where the condition $g/r_0 = E$ is satisfied, a singularity is induced by the localized zero-energy states.}
\label{Coulomb}
\end{center}
\end{figure}

\begin{figure}[t]
\begin{center}
\includegraphics[width=0.45\textwidth]{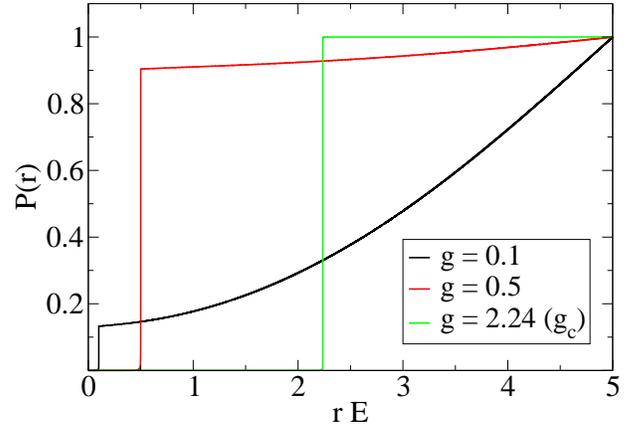}
\caption{Probability of finding one electron within a distance $r$ from the other electron, $P(r) = \sum_i \int_0^{r} d^2 r |\phi_i({\bf r})|^2$, for $K = 0$, $l = 1$, and various values of the dimensionless Coulomb coupling constant $g$, in the important case where a classical turning point exists ($r E = g$). Here, the influence of zero-energy states translates into (i) a suppression of the electron density for $r < r_0$, and (ii) a tendency to concentrate the probability near $r = r_0$ as the critical coupling $g_c$ is approached. The results are normalized to their value at $r E = 5$. }
\label{Prob}
\end{center}
\end{figure}

 The Coulomb problem still presents other peculiarities. In Fig. \ref{Coulomb},  a numerical
 estimate of the radial wave functions is shown for $l = 1$ and two different
 signs of the interaction below the critical value. The main feature in this solution concerns again zero-energy states: when
 the condition $g/r_0 = E$ is fulfilled, zero-energy states must be taken
 into account and become responsible for singularities in the first and third components of the wave function when $l \neq 0$, as shown above on general grounds. In the Coulomb case, this effect has remarkable consequences, such as a drastic suppression of the probability of finding the particle in $r < r_0$, and a tendency to increasingly localize the radial wave function near $r = r_0$ when the critical point $g_c$ is approached from below (see Fig. \ref{Prob}). We find numerically that for $g = g_c$ the relative wave function becomes effectively localized near $r = r_0$.

\begin{figure}[t]
\begin{center}
\includegraphics[width=0.45\textwidth]{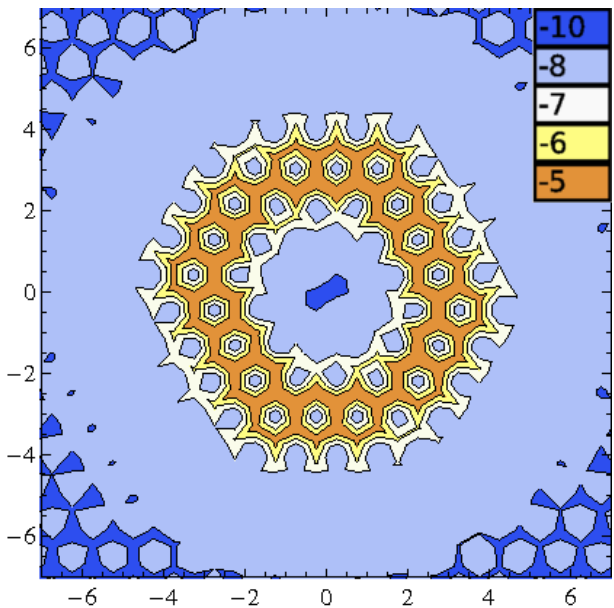}
\caption{Logarithmic density distribution of the two-particle wave function interacting with a Coulomb
potential (see text for details), for $K = 0$. One of the particles is assumed to be placed at the center of the square. The equations are discretized
using the Sutherland lattice, and the energy range of the integrated states is chosen such that they cover the region where the condition $g = r_0 E$ is fulfilled. As expected, the results show a clear concentration of the density at this point $r_0$. The short-length features of the density plot reflect the underlying lattice structure.}
\label{coulomb_2D}
\end{center}
\end{figure}

These results are similar to those obtained for a single particle in the Sutherland lattice \cite{S86} with a Coulomb potential.  As shown in Fig. \ref{coulomb_2D}, we use a
$30 \times 30$ lattice with the structure of Ref. \onlinecite{S86} (see Fig. \ref{Sutherland}) and
periodic boundary conditions.
The potential is $v (r) = v_0 e^{-r/r_d} / \sqrt{r^2+r_1^2}$, with
$v_0 = t > 0, r_d = 20 a$ and $r_1 = 0.5 a$, where $t$ is the hopping,
and $a$ is the distance between nearest neighbor equivalent atoms.
The states considered to construct the density plots are in the range of energies $0.25 t\le E \le 0.35t$. Since they are not eigenstates of the Hamiltonian, they are expected to contain several angular channels. However, as shown in Fig. \ref{coulomb_2D}, this energy spread is sufficient to find an enhancement
of the density in the region near $E = v (r)$. When states in the range $E < 0$ are considered, the density shows a delocalized distribution. In the plots, notice that there are details coming from the underlying lattice structure that are not relevant for our discussion, since we focus on the continuum limit.

In case of considering a small mass in the prob-
lem, the two most prominent features of the Coulomb
two-body problem, i.e., the instability above a critical coupling and the influence of zero-energy states, are not essentially altered.
The Coulomb instability has its origin in the
short-distances behaviour of the wave function,
where mass terms are subleading. These terms are also
 subleading in the expasion around
the classical turning point r0, thus not helping to prevent the appearance of non-analyticities caused by
 zero-energy states which also exist for nonzero mass.

\section{Extension to finite $K$}

The most salient features we have found in the problem of two
interacting  particles are, so far, the influence of zero-energy
states and the appearance of instabilities for the Coulomb
potential. Next we ponder to what extent those results still
apply for the general case of nonzero center-of-mass momentum.

Zero-energy states are investigated by taking $v(r) = E$ in the
eigenvalue problem (\ref{Full_H}). Inspection of the resulting
Hamiltonian reveals that it separates into two independent sectors.
Hence, the Hilbert space of solutions is two-dimensional, with the
general solution for the zero-energy states reading now
\begin{eqnarray}
\psi(r,\phi) = A_1 \left[ \begin{array}{c} r^\alpha \sin^\alpha
(\theta_K - \phi) \\ 0 \\ 0 \\ - r^\alpha \sin^\alpha(\theta_K -
\phi) \end{array} \right] \nonumber \\ + A_2 \left[ \begin{array}{c}
0 \\ \frac{K}{2} r^{\beta} \sin^\beta(\theta_K - \phi) \\
r^{\beta-1} \sin^{\beta-1}(\theta_K - \phi) \\ 0 \end{array} \right]
\, ,
\end{eqnarray}
where $A_1$ and $A_2$, as well as $\alpha$ and
$\beta$, can take arbitrary values. A similar analysis to
that of $K = 0$ can be performed here. For the case of a step potential, they translate
into a change in the matching conditions, since the
introduction of two new parameters ($B_1$ and $B_2$) changes the continuity of the
wave function:
\begin{eqnarray}
\Delta \psi_1(r_0,\phi) &=& B_1(\phi, K, \theta_K)  \nonumber \\
\Delta \psi_2(r_0, \phi) &=& B_2(\phi, K, \theta_K) \nonumber  \\
\Delta \psi_3(r_0, \phi) &=& 0  \nonumber  \\
\Delta \psi_4(r_0, \phi) &=& - e^{2 i \theta_K} B_1(\phi, K, \theta_K)
\end{eqnarray}
Linearization of the equations close to the point where $\varepsilon(r) = 0$ (the classical turning point) shows that, even for smooth potentials, singularities in the relative wave functions arise. Still, they are square integrable and give a finite contribution to the probability.

As for the Coulomb instability, its existence can be probed by
checking  the short-distance limit of the full Hamiltonian given in
(\ref{Full_H}), for the case $v(r) = g/r$. It is not difficult to
see that the small $r$ limit is controlled by $K$-independent terms. Thus, for $r \rightarrow 0$ we recover the $K = 0$ limit, where an instability has already been identified. Hence, we expect this instability to be a general feature of the Coulomb problem.

\section{Relevance to many-body phenomena}

As mentioned in the Introduction, it can be expected  on general
grounds that the two-body problem with Coulomb interactions
provides information on the more complicated many-body problem
in graphene. This is especially relevant in the strong-coupling
regime, as several works in the literature suggest the
possibility of a new insulating phase above a certain critical
coupling, where electron and holes would bind forming excitons and
opening up a gap.

It has been pointed out in the literature \cite{GGG09, Wang09},
that the breakdown of the Dirac vacuum in the attractive Coulomb
impurity problem could be related to such a formation of excitons in
graphene for strong enough coupling. We have seen in this article that this behavior is also present in the two-body problem, which should be
even more relevant to the many-body physics

In order to understand the connection, we must notice that in
principle  the problem of an interacting electron and hole can be
mapped into that of two attractive electrons, with similar symmetry
properties. However, like for the Coulomb impurity problem, a
more rigorous mapping would be performed by considering the
existence of the Dirac sea, which constraints,
by Pauli's principle, the states accessible to the electron-hole
pair, in analogy to the Cooper problem \cite{Schrieffer99}. Such a
treatment, however, would require to work in a different basis for which the analytical results obtained in this paper do not hold
\footnote{J. Sabio, F. Sols and F. Guinea, to be published}, and is thus beyond the scope of this paper.

We note in this regard that the breakdown  of
the Dirac vacuum in the two-body problem  could be a signature of the
excitonic instability in the many-body system. As we have already
shown, the critical value for which the breakdown occurs depends on
the scattering channel. For the most
symmetric one, the $s$-wave, we find $g_c = 1$, which should be compared to the critical values obtained for the Coulomb impurity problem, $g_c^{{\rm CI}} = 0.5$. \cite{PNN07,
Shy07, Shy07y2, N07} For higher
angular-momentum channels the critical couplings increase. As an example,
$g_c = 2.24$ for $l = 1$. However, at low energies those higher angular momenta
are usually less important.  Hence, the $s$-wave
critical coupling should provide an educated guess of the corresponding value for the expected
many-body instability. Remarkably, the critical values obtained so far in the theoretical literature are close to the value predicted here for the two-body problem. Monte Carlo calculations in the lattice give a critical coupling $g_c^{{\rm MC1}} \simeq 1.11$ \cite{DL09, DL09b} and $g_c^{{\rm MC2}} \simeq 1.66$, \cite{AHS09} depending on the model used to simulate graphene electrons. Renormalization Group calculations yield $g_c^{{\rm RG}} \simeq 0.833$. \cite{Vaf08} Finally, a variational approach to the excitonic condensate has been recently reported to show a transition above the critical coupling $g_c^{{\rm var}} \simeq 1.13$. \cite{Khveshchenko09} The two-body problem with Coulomb interactions of strength above the critical coupling has not been addressed in this paper. As mentioned above, from the study of the Coulomb impurity problem it can be expected that the instability, which produces a reorganization of the Dirac vacuum, leads to a non-linear screening of the interactions. \cite{Shy07y2} This effect should be analyzed carefully in order to establish its connection with a possible formation of excitons, where it could happen that still the presence of the Dirac sea as a constraint is necessary in order to produce a bound state. After all, the results presented in this paper do not shed sufficient light on the consequences of the actual many-body instability.

Regarding the spin degree-of-freedom, as discussed in section II, if
both electrons belong to the same valley the $s$-wave channel would
correspond to a triplet state in the spin sector. This fact may be
highly relevant for the study of the excitonic instability in the
presence of an external magnetic field.

There is a second aspect of the two-body problem in graphene that could have consequences on the more complicated many-body problem: the influence of zero-energy states for $l \neq 0$ angular momentum channels. As we have seen, in those cases where the kinetic energy vanishes at some point (positive total energy and repulsive potential, or negative total energy and atractive potential), zero-energy states induce singularities in the wave function which translate into an increasing probability of finding the particle at the classical turning point $r_0$, when the critical coupling $g_c$ is approached.

Let us discuss the role of the carrier density in this scenario. We only consider couplings below the critical one in case of Coulomb interactions, since we expect a description in terms of linear screening to be applicable. In doped
graphene, electrons at the Fermi surface have an energy $E_{F}=v_{F}k_{F}
= v_F ( 4 \pi n / N_s)^{1/2}$, where $n$ is the electron density in the upper cone and $N_s = 4$ the valley and spin degeneracy. This
defines a classical return distance $r_{0}\equiv g/E_{F}\propto n^{-1/2}$
for the Fermi surface electrons at which density correlation should
peak. On the other hand, the static screening of the Coulomb interaction in
doped graphene is characterized by the Thomas-Fermi (TF) screening length $%
\lambda _{\mathrm{TF}} = g^{-2}(4 \pi n N_s)^{-1/2}$,\cite{Ando06,Wun06,Hwa07} which shows a similar density dependence, namely, $%
\lambda _{\mathrm{TF}}\propto n^{-1/2}$.
Thus the ratio between the classical return and screening distances is
independent of the density: $r_{0}/\lambda _{\mathrm{TF}}=N_{s}g^{2}$. For
many cases we expect $r_{0}/\lambda _{\mathrm{TF}}>1$, which places $r_{0}$
beyond the screening length, i.e. where the bare Coulomb interaction, for which $%
r_{0}$ has been calculated, does not hold. Naively this might invalidate the
physics associated to zero-energy states, which is expected to occur at $%
r=r_{0}$. However, it is easy to note that the density correlation peaks
have to be a robust feature of the many-body problem. 

We have seen that
zero-energy states intervene at the point where $v(r)=E$. It is quite
reasonable to assume that, in a many-body context, that condition must be
replaced by $v_{\mathrm{scr}}(r_{0})=E_{F}$, which defines the classical return distance $r_{0}$ for the electron gas if $v_{\mathrm{scr}}(r)$ is the
screened Coulomb interaction potential. Within the TF
approximation, the screened potential has the form \cite{Ando82,GV05}  $v_{\mathrm{scr}}(r_{0})=(e^{2}/r)F(r/\lambda _{\mathrm{%
TF}}),$ where $F(x)$ is a monotonically decreasing function satisfying $%
F(x)\simeq 1$ for $x\ll 1$ and $F(x)\sim x^{-2}$ for $x\gg 1$. Dimensional
analysis shows that the dressed $r_{0}$ also scales like $n^{-1/2}$, which
suggests that zero-energy states play a role even in the presence of screening.

\section{Conclusions}

The study of two interacting Dirac electrons in
graphene has led us to unveil intriguing properties of charge carriers in this novel material. On the one hand, due to the chiral nature
of the low-energy electrons, the center-of-mass and
relative coordinates are coupled even in the presence of central
potentials. This precludes a simple decomposition in terms of an
effective one-body scattering problem. However, in the case of zero total momentum, the two-body problem can be mapped into that of a single particle in the Sutherland lattice \cite{S86}.

Zero-energy states turn out to play a pivotal role in the scattering processes, changing the
matching conditions in the simple case of a step potential, and introducing singularities in the wave function for general potentials, including Coulomb.

The case of Coulomb interaction is most relevant for the analysis of strong coupling instabilities. The reason is, electrons in weakly doped graphene have poor screening properties that, unlike in the conventional two-dimensional electron gas \cite{GGV99, GFM08}, are expected to
preserve the long-range tail of this potential.
Although the problem cannot be exactly mapped into the Coulomb
impurity problem, widely studied in the literature, it still shows
similar features such as the existence of a critical coupling above
which wave functions become ill-defined, a likely signature of
the Dirac vacuum breakdown. In a many-body context, this
could signal the formation of a new insulating phase characterized by
electron-hole pairing. An analysis of the $s$-wave scattering channel
for $K = 0$ gives a critical
coupling for the instability of $g_c = 1$, in rather good agreement with the critical values obtained in theoretical studies of the full many-body problem. We may also note that, due to the
symmetry properties of the wave functions, this $l=0$ channel
is accompanied by a spin triplet state if, as assumed throughout this paper, both
particles belong to the same valley.

The Coulomb potential shows other interesting properties. We have shown that the effect of zero-energy states in $l \neq 0$ angular channels is that of partially localizing the electron density near the classical turning point. The degree of localization increases dramatically as the critical coupling $g_c$ is approached. Quite generally, we may expect the novel features found in the two-body problem to have wide ranging implications on the many-body problem in graphene lattices.

\section*{Acknowledgements.}
We are grateful to Simone Fratini and Anthony J.  Leggett for
helpful discussions. We acknowledge financial support by MEC (Spain)
through grants FIS2007-65723, FIS2008-00124 and CONSOLIDER
CSD2007-00010, and by the Comunidad de Madrid through CITECNOMIK.
J.S. also wants to acknowledge the I3P Program from the CSIC for
funding.

\section*{APPENDIX A: Derivation of the matching conditions}

As mentioned in the main text (section IV), the solutions located at
the point $v(r)=E$ can induce discontinuities in the wave function.
Physically, this can be understood in terms of localized states that live in
this region and which are built from the complete set of polynomical solutions
given in (\ref{localized}). Let us develop the argument in detail.

For greater clarity, we may modify the step potential near the point $r_{0}$
where $v(r_{0})=E$ in such a way that $v(r)=E$ [i.e. $\varepsilon (r)=0$] in
the vicinity of $r_{0}$, namely, for $r_{0}-\delta <r<r_{0}+\delta $, where
at the end of the calculation $\delta \rightarrow 0$. \ From Eqs. (\ref%
{phi-2}) and (\ref{par-phi-2}) in the main text, it follows that $\partial
_{r}\phi _{2}=0$ and, if $l\neq 0$, $\phi _{2}=0$. On the other hand, Eq. (%
\ref{system-2}) becomes, in that small interval,%
\begin{equation}
r\partial _{r}(\phi _{1}-\phi _{3})=(l-1)\phi _{1}+(l+1)\phi _{3}~.
\label{diffzero-1}
\end{equation}

The existence of zero-energy states [see Eq. (\ref{localized})] allows us to
introduce functions of arbitrarily high slope in the small interval of
length $\delta $. We adopt the simplest ans\"{a}tze for the two components:
\begin{eqnarray}
\phi _{1}(r) &=&a+b(r-r_{0}) \\
\phi _{3}(r) &=&c+d(r-r_{0})\,.
\end{eqnarray}%
In the slightly modified potential, both components must be continuous
everywhere, so we may impose
\begin{eqnarray}
\phi _{1}(r_{0}\pm \delta ) &=&a\pm b\delta \equiv \phi _{1}^{I,II}(r_{0}) \\
\phi _{3}(r_{0}\pm \delta ) &=&c\pm d\delta \equiv \phi _{3}^{I,II}(r_{0})\,.
\end{eqnarray}%
As a result, if $\Delta \phi _{i}\equiv \phi _{i}^{II}(r_{0})-\phi
_{i}^{I}(r_{0})$,
\[
b=\Delta \phi _{1}/2\delta ,\,~~~~d=\Delta \phi _{3}/2\delta \,.
\]%
If we allow for nonzero discontinuities, $\Delta \phi _{i}\neq 0$,
we conclude that, for $\delta \rightarrow 0$, both $\phi
_{1}^{\prime }=b $ and $\phi _{3}^{\prime }=d $ tend
to infinity in magnitude. Thus, Eq. (\ref{diffzero-1}) can be
approximated as
\begin{equation}
\partial _{r}\phi _{1}(r)=\partial _{r}\phi _{3}(r)\,,
\end{equation}%
i.e. $b=d$ and thus%
\begin{equation}
\Delta \phi _{1}=\Delta \phi _{3}\equiv -2C  \label{Del-phi13}
\end{equation}

The upshot is that, thanks to the existence of zero-energy states in the
immediate vicinity of $r_0$, a new parameter emerges that allows for a
discontinuity in the components $\phi_1$ and $\phi_3$. The parameter $C$ is
thus adjusted to render the matching problem well determined.

Interestingly, if one were to perform a similar analysis to the one-body
problem of a step potential impurity, one would introduce a similar ansatz
for the (only existing) two components of the problem. Zero-energy states
could in principle also play a role in the vicinity of the point analogous
to $r_{0}$. However, we find that a linear ansatz similar to that considered
above would lead to a zero slope. In other words, even allowing for the existence of
zero-energy states, the wave function remains continuous at all points,
including $r_{0}$. We could say that zero-energy states do not intervene
because they are not necessary, and this is so because, unlike in the
two-body problem, the matching problem is well defined from the start.

Once we have taken $\delta \rightarrow 0$ and accepted that the abrupt
change of sign of $\varepsilon (r)$ at $r=r_{0}$ leads to identical
discontinuities in $\phi _{1}$ and $\phi _{3}$ while keeping $\phi _{2}$
continuous, we may derive, from the general relations in section III.B, a
few more conclusions on the behavior of the solutions around the step.

From the fact that $\varepsilon $ and $\phi _{1},\phi _{3}$ are bounded, it
follows from Eqs. (\ref{phi-2}) and (\ref{par-phi-2}) that $\phi _{2}$ and $%
\partial _{r}\phi _{2}$ are also bounded. If we integrate Eq. (\ref{system-1}%
) in an infintesimally small region around the step, we conclude%
\begin{equation}
\Delta (\varepsilon (\phi _{1}+\phi _{3}))=0~,  \label{Del-eps-phi13}
\end{equation}%
where $\Delta $ means total variation across the abrupt step. If we combine
this result with Eq. (\ref{Del-phi13}), we conclude that the common
discontuity of $\phi _{1}$ and $\phi _{3}$ is directly determined by the
step discontinuity in the potential ($\Delta \varepsilon =-v_{0}$).
Therefore Eq. (\ref{Del-eps-phi13}) implicitly yields the discontinuity $C$ which is needed to allow $\phi _{2}$ to be
continuous. Specifically, we obtain%
\begin{equation}
C=\frac{1}{4}\left( 1-\frac{|\varepsilon ^{I}|}{\varepsilon ^{II}}\right)
\left( \phi _{1}^{I}+\phi _{3}^{I}\right) ~,
\label{C_eq}
\end{equation}%
where $\varepsilon ^{I}=E-v_{0}<0$ and $\varepsilon ^{II}=E>0$.

From Eq. (\ref{par-phi-2}) it follows that $\partial _{r}\phi _{2}$
experiences a discontinuity across the step which closely follows the
discontinuity of $\varepsilon (r)$, given that $\phi _{1}-\phi _{3}$ is
continuous. We also note from Eq. (\ref{phi-2}) that, for $l\neq 0$, $\phi_2$ goes
quickly through zero as $\varepsilon (r)$ becomes 0 at $r=r_{0}$. However, \
it recovers quickly from that sharp dip to become globally continuous across
the step, as can be inferred from (\ref{phi-2}) and (\ref{Del-eps-phi13}).
By contrast, when $l=0$, $\phi _{2}$ remains strictly continuous across the
step.

\begin{widetext}
\section*{APPENDIX B: Analytical solution of the short-range
interacting problem for zero center-of-mass momentum}

We start from the scattering problem sketched in Fig. \ref{step}.
The  two electron problem is written in terms of an effective
single-electron radial equation in the case $K = 0$. The energy of
the incident pair is $ E < v_0$. For $r < r_0$, only solutions
non-singular at the origin are valid, while for $r > r_0$, a general solution is a
linear combination of incoming and outgoing wave functions. Hence we
have, up to a normalization constant [see Eqs.
(\ref{besselJ})-(\ref{E-solutions})],
\begin{eqnarray}
\phi^I_l = \left[ \begin{array}{c} \frac{1}{2} J_{l-1}(k_I r)  \\ -
J_{l}(k_I r) \\ \frac{1}{2} J_{l+1} (k_I r) \end{array} \right]
\nonumber \\ \phi^{II}_l =  A \left[ \begin{array}{c} \frac{1}{2}
J_{l-1}(k_{II} r) \\ J_{l}(k_{II} r) \\ \frac{1}{2} J_{l+1} (k_{II}
r) \end{array} \right] + B  \left[ \begin{array}{c} \frac{1}{2}
Y_{l-1}(k_{II} r) \\ Y_{l}(k_{II} r) \\ \frac{1}{2} Y_{l+1} (k_{II}
r) \end{array} \right] \label{phi-AB}
\end{eqnarray}
where the coefficients of the wave function are those of positive
energy for region I and those of negative energy for region II. Moreover, in
Eq. (\ref{phi-AB}), $k_I = (v_0-E)/2$ and $k_{II} = E/2$.

As already seen in Appendix A, both solutions must be matched  at $r
= r_0$ with a matching condition that includes an arbitrary
coefficient, say $C$, to be adjusted. The system of equations reads
now:
\begin{eqnarray}
\left[ \begin{array}{ccc} \frac{1}{2} J_{l-1}(k_{II} r_0) &
\frac{1}{2} Y_{l-1}(k_{II} r_0) & 2 \\  J_{l}(k_{II} r_0) &
Y_{l}(k_{II} r_0) & 0 \\ \frac{1}{2} J_{l+1}(k_{II} r_0) &
\frac{1}{2} Y_{l+1}(k_{II} r_0)  & 2 \end{array} \right] \left[
\begin{array}{c} A \\ B \\ C \end{array} \right] = \left[
\begin{array}{c} \frac{1}{2} J_{l-1} (k_I r_0) \\ - J_{l}(k_I r_0)
\\ \frac{1}{2} J_{l+1}(k_I r_0) \end{array} \right]\, ,
\end{eqnarray}
which can be solved by using Cramer's method. Invoking
Bessel function properties, the coefficients are found to be
\begin{eqnarray}
A = -\frac{\pi r_0}{2}\left[J_{l}(k_I r_0) \frac{d}{d r_0} Y_{l}(k_{II} r_0)
+ \frac{k_{II}}{k_I} Y_{l}(k_{II} r_0) \frac{d}{d r_0} J_{l}(k_I r_0)\right]\\
B =  \frac{\pi r_0}{2}\left[J_{l}(k_I r_0) \frac{d}{d r_0} J_{l}(k_{II} r_0)
+ \frac{k_{II}}{k_I} J_{l}(k_{II} r_0) \frac{d}{d r_0} J_{l}(k_I r_0)\right]\\
C =  \frac{l}{4} J_{l}(k_I r_0) \left(\frac{1}{k_{II} r_0} + \frac{1}{k_I r_0}\right)
\end{eqnarray}
These analytical expressions reproduce the numerical results
obtained  by discretizing the differential equations, including the
magnitude of the jump, $-2C$, and Eq. (\ref{C_eq}) from Appendix A.

\section*{APPENDIX C: Analytical solution of the s-wave channel
for the Coulomb interacting problem}

We start from the system of differential equations  given in
(\ref{Coulomb_system}). The $s$-wave channel corresponds to $l = 0$.
In this case, $\hat{\phi}_1 = - \hat{\phi}_3$, and the system
reduces to one of only two components, with a structure resembling that of the
Coulomb impurity problem. We define:
\begin{eqnarray}
Q_1 = \hat{\phi_1} - \frac{i}{2} \hat{\phi_2}\\
Q_2 = \hat{\phi_1} + \frac{i}{2} \hat{\phi_2}
\end{eqnarray}
which fulfill the following coupled differential equations:
\begin{eqnarray}
(\rho \partial_{\rho} + \gamma - i\frac{g}{2})Q_1 + \frac{Q_2}{2} = 0\\
(\rho \partial_{\rho} - \rho + \gamma + i\frac{g}{2})Q_2 + \frac{Q_1}{2} = 0\\
\label{eq_Q}
\end{eqnarray}
The solutions are given by Kummer functions \cite{Abramovich65}:
\begin{eqnarray}
Q_1 = C_1 {\mathcal F}(\gamma-i\frac{g}{2}, 2\gamma+ 1; \rho)\\
Q_2 = C_2 {\mathcal F}(\gamma + 1-i\frac{g}{2}, 2\gamma+ 1; \rho)\\
\end{eqnarray}
By using the property ${\mathcal F}(a,b;0) = 1$ and the  limit $\rho
\rightarrow 0$ of the system of equations (\ref{eq_Q}), we obtain the
ratio:
\begin{equation}
c_{21} \equiv \frac{C_2}{C_1} = - 2(\gamma - ig/2) =  e^{- i
\arctan{\frac{g}{2\gamma}}}
\end{equation}
Hence the solution is:

\begin{eqnarray}
\phi(r) \sim \frac{1}{2} (i E r)^{\gamma-1/2}  e^{-\frac{i E r}{2}}
\left[ \begin{array}{c} {\mathcal F}(\gamma-i\frac{g}{2}, 2\gamma+
1; i E r) + c_{21} {\mathcal F}(\gamma + 1 -i\frac{g}{2}, 2\gamma+
1; i E r)\\ 2 i{\mathcal F}(\gamma-i\frac{g}{2}, 2\gamma+ 1; i E r)
-  2 i c_{21}{\mathcal F}(\gamma + 1 -i\frac{g}{2}, 2\gamma+ 1; i E
r)\\ - {\mathcal F}(\gamma-i\frac{g}{2}, 2\gamma+ 1; i E r) - c_{21}
{\mathcal F}(\gamma + 1 -i\frac{g}{2}, 2\gamma+ 1; i E r)
\end{array} \right] \end{eqnarray}

up to an overall normalization constant that can be  determined by
matching the solution to the $r\rightarrow \infty$ limit.
\end{widetext}

\bibliography{twoDirac}

\end{document}